# Reduced heat flow in light water ($H_2O$) due to heavy water ($D_2O$)

William R. Gorman and James D. Brownridge[a]
State University of New York at Binghamton
Department of Physics, Applied Physics, and Astronomy
Binghamton, NY 13902

The flow of heat, from top to bottom, in a column of light water can be decreased by over 1000% with the addition of heavy water. A column of light water cools from 25°C to 0°C in 11 hours, however, with the addition of heavy water it takes more than 100 hours. There is a concentration dependence where the cooling time increases as the concentration of added $D_2O$ increases, with a near maximum being reached with as little as 2% of $D_2O$ added. This phenomenon will not occur if the water is mixed after the heavy water is added.

Increasing the thermal conductivity of liquid water will increase the rate that heat can be transferred from one source to another.[1-5] Although much effort has been devoted to enhancing thermal conductivity there seems to be little devoted to its reduction. Reducing the thermal conductivity of liquid water will allow heat to remain longer which will increase our ability to store thermal energy in bulk water, producing a major impact on energy conservation.[6] Heavy water was added to light water in incremental volumes such that the concentration increased from 0.023% to about 2%. Heavy water occurs in light water at about 0.015% by molar volume and we used heavy water that was 99.9% pure. We report here only on heavy water added to light water. However, a similar effect was observed when light water was added to heavy water. This phenomenon has also been found to occur when other chemicals were added to water. However, in this study we concentrated our efforts on heavy water ($D_2O$) because of the similarities between it and light water ($H_2O$).[7,8] One of the most dramatic effects of adding $D_2O$ to $H_2O$ is the increase in time that it takes a volume of water to cool, as shown in Fig. 1.

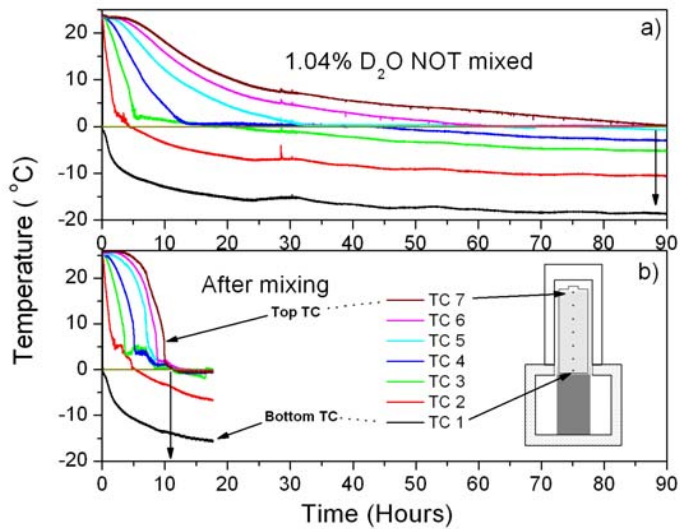

**FIG. 1.** Seven thermocouples were positioned in a one liter bottle as shown as the inset to Fig. 1b. Figure 1b shows how long it took all of the thermocouples reach ≤ 0°C when the water was "pure" $H_2O$. In comparison, Fig. 1a shows the effect of adding 1.04% $D_2O$ just prior to cooling. This figure clearly shows that the time for all of the thermocouples to cool to roughly the same temperature is substantially longer. The only variable is unmixed heavy water distributed throughout the light water.
*a) jdbjdb@binghamton.edu*

This phenomenon may be produced by carefully adding variable amounts of $D_2O$ to $H_2O$ and then cooling the container before the $D_2O$ mixes with the $H_2O$[9-14]. When heat is removed from the bottom of a container, the time that it takes the top of the container to cool to a specific temperature is found to depend on the amount of $D_2O$ added, how much time elapsed following the addition of the $D_2O$ and the temperature of the cooling environment. By keeping the cooling environment and the elapsed time the same; we were able to determine how the concentration of added $D_2O$ changed the effect seen. For our setup, the time it took the top thermocouple in our container to cool to ≤ 0°C as a function of different amounts of added $D_2O$ is shown in Fig. 2; the cooling environment was constant at -21°C.

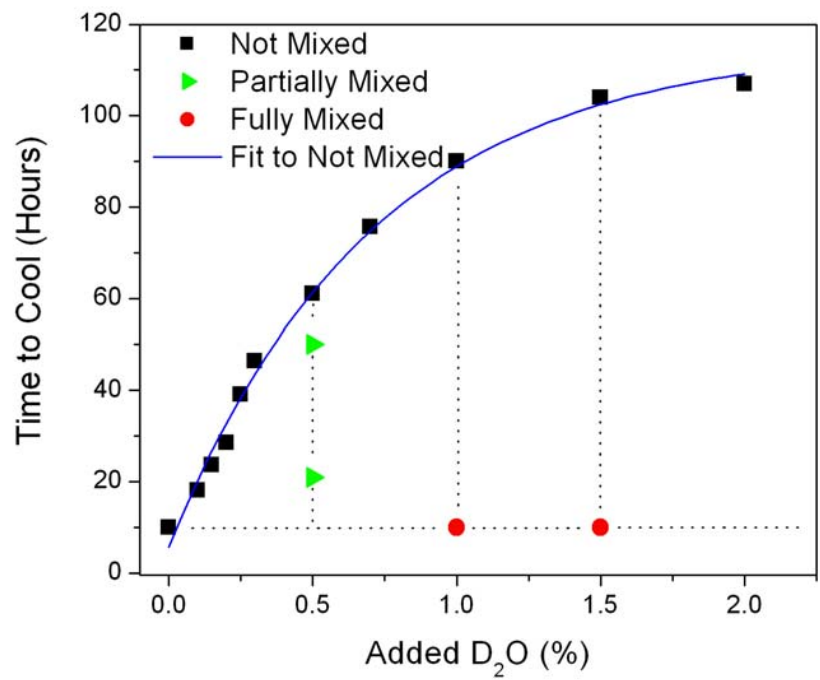

**FIG. 2.** As the concentration of $D_2O$ increases, the time it takes the top thermocouple to reach 0°C also increases up to some limit. The setup is as shown in FIG. 1b. An exponential fit of the Not Mixed data gave a maximum time of 116 ± 4 hours.

If the $D_2O$ is added and then mixed (done by vigorously shaking the bottle), it will cool as if no $D_2O$ was added. This is shown in Fig. 2 by the difference in times it took for the 1.0 % and 1.5 % concentration for mixed and unmixed to cool to 0°C. It was also determined that the time to cool depended upon the amount of mixing as well. When mixing was not complete the time varied as shown for the two points at 0.5% $D_2O$. Here the bottle was gently rotating 180° and 90° to produce the data points at 20 and 50 hours respectively. Experiments indicate that with 0.023% $D_2O$ added to $H_2O$ and left in a controlled environment, it will take over 90 hours to self mix.

Thermal fluctuations had developed during the cooling process of a 110 mL container and were investigated. There were 6 thermocouples placed in the center of the container such that the covered an area less than 1 cm$^2$. Figure 3 shows the magnitude of the thermal fluctuations when 0.46% $D_2O$ is added to $H_2O$ whereas the inset shows the temperature gradient for $H_2O$ with no $D_2O$ added.

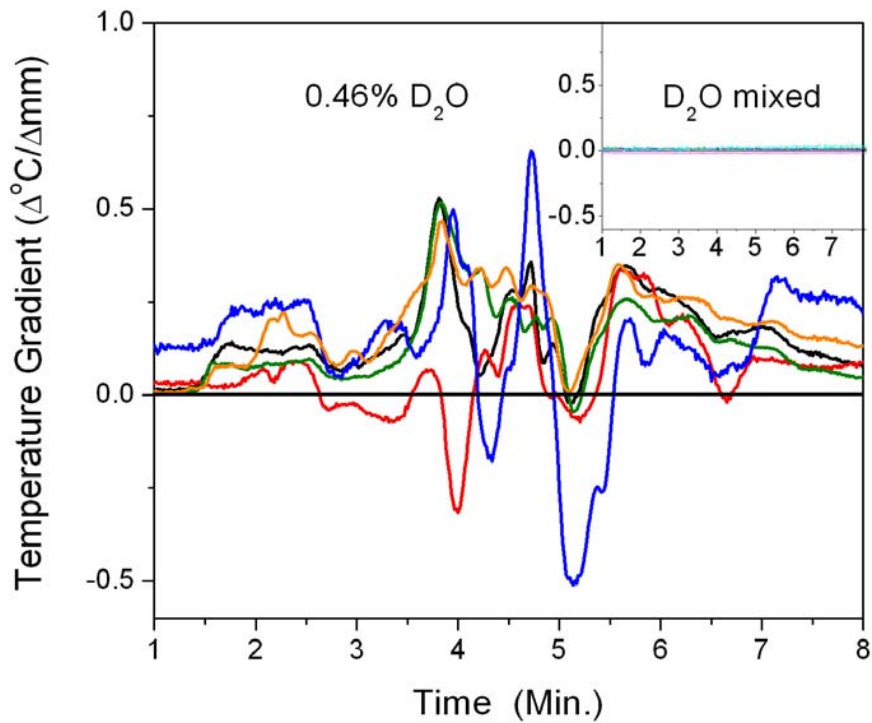

**FIG. 3.** Here we have placed 6 thermocouples horizontally near the center of a glass container holding 110 mL of water. The thermocouples are spread over an area less than 1 cm$^2$. The inset is the graph one would get when the water has no heavy water added or if the heavy water is vigorously mixed.

One of the most important key factors for this effect to set up is that the $D_2O$ must "filter" through $H_2O$ by adding to $D_2O$ to the top of the container. If the $D_2O$ is added to the bottom of the container, it will not move throughout the container and thus the effect will not be seen. Also if the water is agitated, stirred, or mixed in any way, the effect is reduced or totally lost.

We observed that when heavy water was added at a concentration near 0.4%, the isothermal line that naturally develops in still water as it cools through 4°C did not develop.[15] The isothermal line that naturally occurs in still water is seen as a sudden drop in temperature and since an isothermal line moving up the column of water is a mode of heat transfer, then preventing this line from forming is a method to reduce heat transfer. We also observed that with increasing concentrations of unmixed $D_2O$ it takes longer for water at the center of a container to begin to cool after the container was placed in a cold environment. This then suggests that the unmixed $D_2O$ is somehow blocking one mode of heat transfer from the water to its environment. This could explain why it takes longer for water with added unmixed $D_2O$ to cool. Furthermore, the thermal fluctuations that were seen during the cooling process indicate that there was a resistance to the transfer of heat from the position of one thermocouple to the next during cooling, as seen in Fig. 3.

We believe that dissimilar water clusters hold the key to understanding the origin of this phenomenon[12]. After adding heavy water to light water there will initially be two types of water clusters. One is $H_2O$ dominated and the other is $D_2O$ dominated. The $H_2O$ and $D_2O$ dominated clusters both consist of a collection of $H_2O$, $D_2O$, DHO and HDO molecules because neither is 100% pure. In these experiments, the $H_2O$ dominated clusters initially consisted of 99.985% $H_2O$ molecules with the $D_2O$ dominated clusters initially consisting of 99.9% $D_2O$ molecules. As long as the clusters remain intact, the dominant molecule in each cluster governs its response to a change in the thermal environment. The size and shape of the clusters may vary and they are able to break apart and reform at rates that depending on the temperature and condition of the physical environment, i.e. shaking and stirring[12,16,17]. Eventually, all clusters break apart and reform which results in a homogeneous mixture of $H_2O$ dominated clusters and the effect disappears. This is clearly shown in Fig. 2 after we broke apart the $D_2O$ dominated clusters by shaking the container and they then reformed as homogeneous mixture of $H_2O$ dominated clusters. These clusters are also believed to be responsible for the thermal oscillations that are present. The clusters cause a resistance to the transfer of heat and thus cause an increase in temperature. As the resistance will build up enough to over come the cluster and allow the heat to be transferred which will then decrease the temperature. This is the thermal fluctuations that are seen at work.

Heat flow in light water can be reduced up to 1000% with the simple addition of a small percent of $D_2O$. We believe it is the heavy water dominated clusters lodged between light water dominated clusters that cause the delay in heat transfer. We have shown that there is a concentration dependence which causes the magnitude of this effect to increase as the concentration increases. A unique property of this effect is the fact that the $D_2O$ has to filter through the $H_2O$ for the effect to occur and any method of mixing will reduce or completely remove the effect. This effect occurs as long as the heavy water clusters stay intact in the light water environment which lasted up to 90 hours in our experiments.

**Acknowledgement**

We would like to thank Binghamton University's Department of Physics, Applied Physics, and Astronomy for its support.

# SUPPLEMENTAL INFORMATION

## Methods and experimental setups

## Cooling from the bottom experiments

This is the setup for cooling from the bottom which produced the data presented in Figure 1 in the main article. For this series of experiments 960 mL of ordinary deionized water was added to a one liter plastic bottle with seven type K thermocouples in it. The tips of the thermocouples were evenly positioned axially in the center of the bottle. The top thermocouple was just under the surface of the water and the bottom thermocouple was in contact with the bottom of the bottle. To begin the cooling process, the bottom of the bottle was inserted ~2 cm into a cold plate which contained a glycol solution at -21$^o$C. The glycol solution ensures good thermal contact. The bottle was covered with a vacuum flask to ensure minimum lost of heat through the sides during cooling. The objective was for all heat to flow from the top to bottom so that no convection currents would develop during cooling. Just prior to placing the bottle on the cold plate $D_2O$ or ordinary deionized $H_2O$ (as a control) was very carefully added to the top with a dropper, drop by drop from a height of about 0.5 cm above the surface of the water. The temperature of each thermocouple was recorded once per second beginning several minutes before the addition of $D_2O$ or ordinary deionizer $H_2O$ and continued until the top thermocouple reached 0$^o$C or lower. We use the term

"ordinary" to indicate that we are talking about water with deuterium at the naturally occurring concentration of 0.014 - 0.015 %.

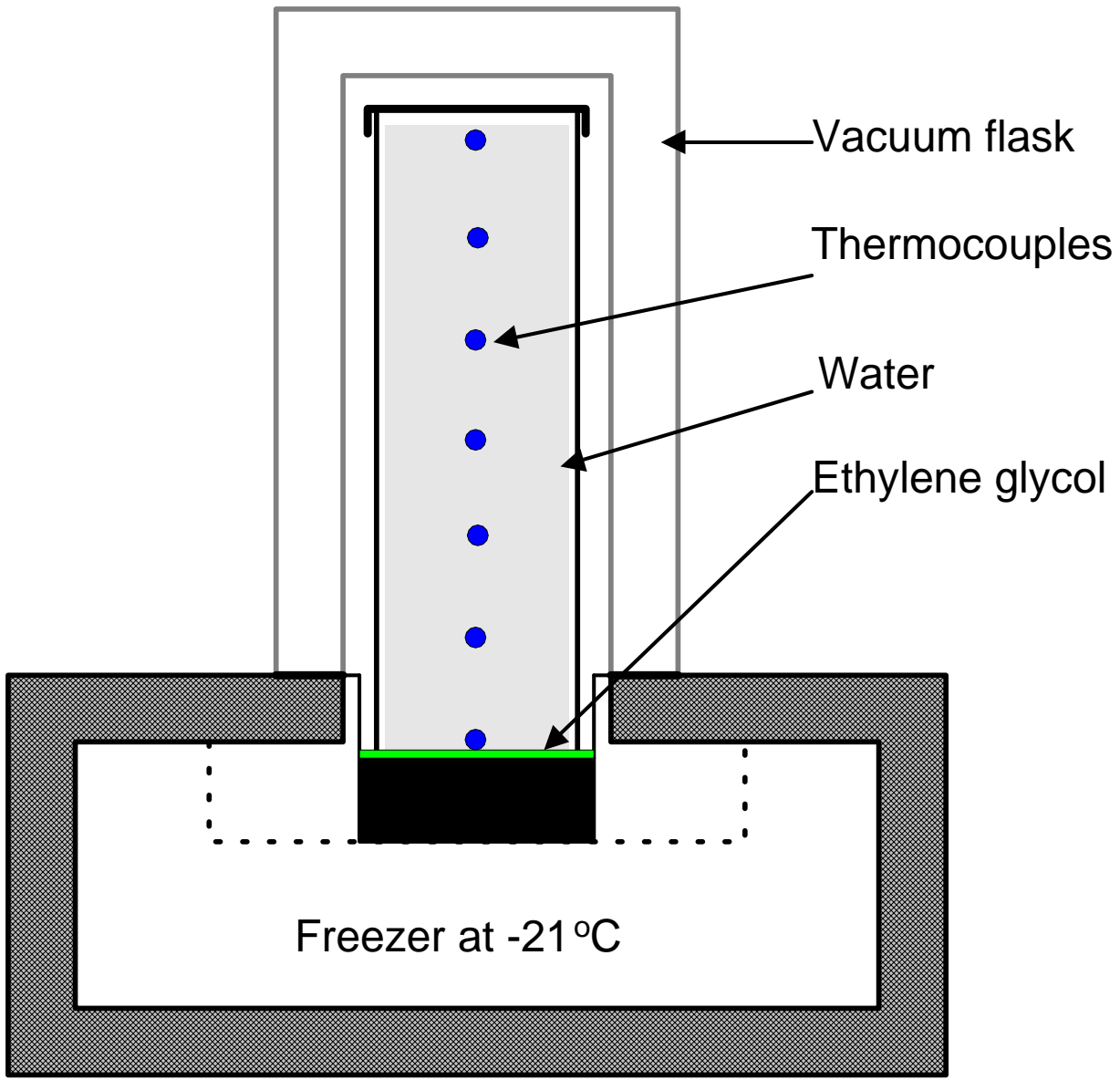

**Fig. 1S.** The schematic of the setup used to remove heat only from the bottom of a container of water. This arrangement permits the removal of heat without the introduction on convection currents in the water.

# Cooling from all surfaces

In a second setup (below), six thermocouples were placed in the center of a bottle containing 110 mL of water. The object here was to remove heat from all surfaces of the bottle and do it quickly by inserting the bottle into an ice bath. Figs. 2S-6S shows the temperature gradients that develop when the container was cooled. Here we show that as the concentration of added $D_2O$ is increased, the thermal fluctuations increase. **The thermal gradient that develops in ordinary water, shown in Fig. 2S, is an isothermal line, created in part by convection, that is moving past the thermocouples as the water passes through its point of maximum density at ~4°C. Notice that after the addition of $D_2O$, the signature of the isothermal line previously seen in Fig. 2S is suppressed, as shown in Fig. 3S-6S. The signature of an isothermal line, if it is produced, will be observed only as the water cools through 4°C.**

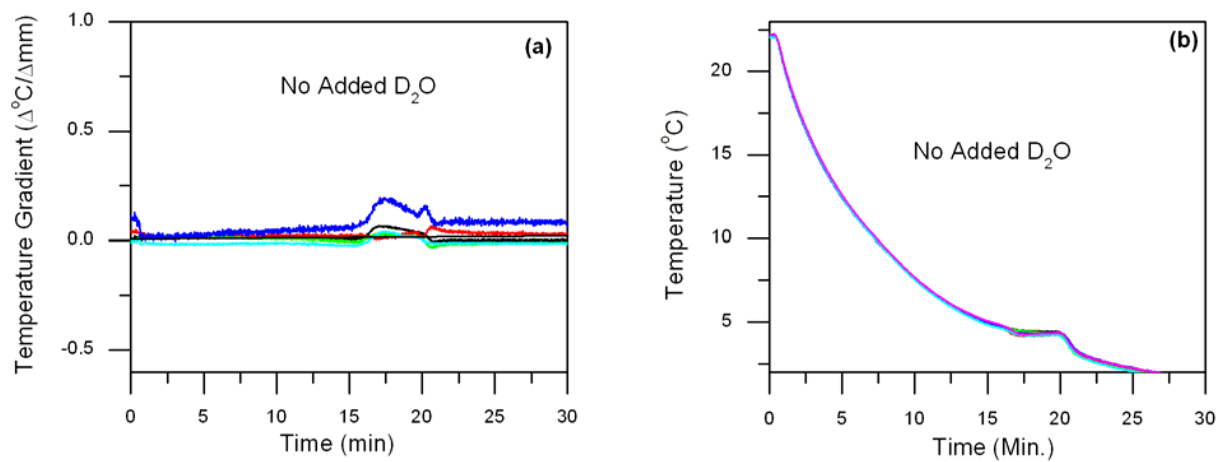

**Fig. 2S.** Here we show a graph **(a)** of the temperature gradient produced in the center of the bottle when it is filled with regular water with no added $D_2O$. The gradient at T = ~0 was always seen when we inserted the bottle into the ice bath. The gradient at T > ~15 min is the 4 °C isothermal line moving up the bottle. Graph **(b)** is the cooling curve from which **(a)** was derived.

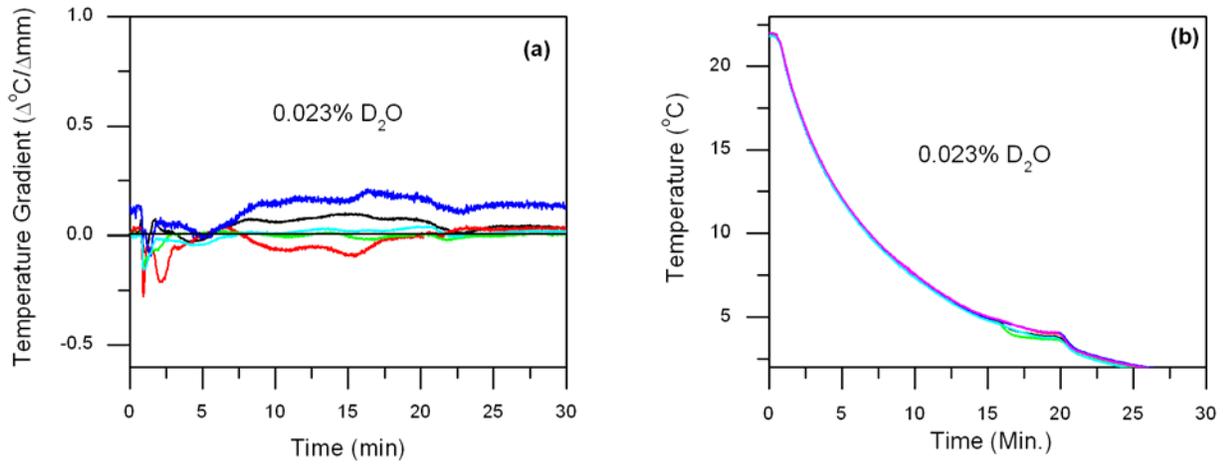

**Fig. 3S**  Here we have added only about twice the amount of $D_2O$ that naturally occurs in water. However, it is not homogenously mixed with the water as is the natural $D_2O$ is. The effect is quite dramatic. Graph **(b)** is the cooling curve from which **(a)** was derived.

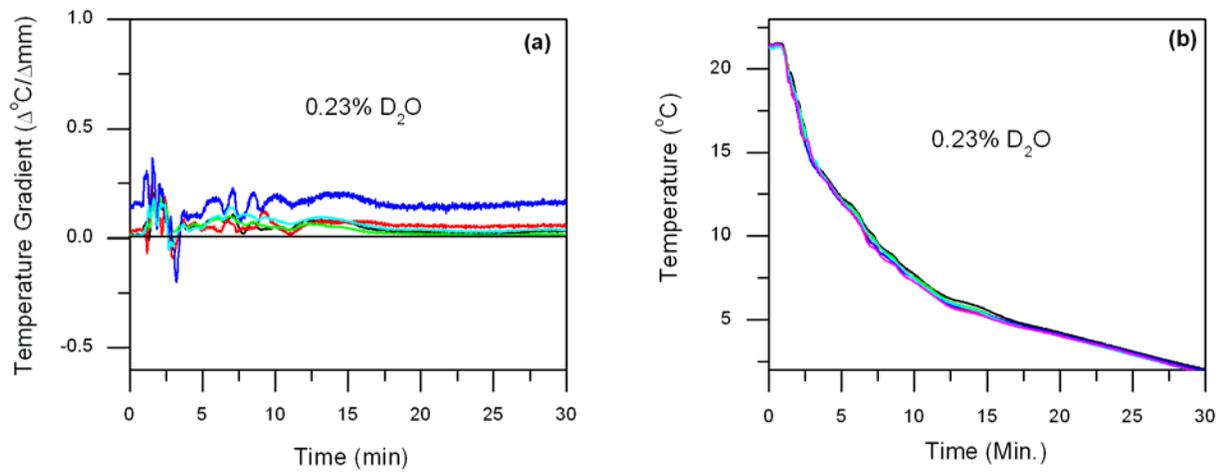

**Fig. 4S**  Here we add 0.23 %. Graph **(b)** is the cooling curve from which **(a)** was derived.

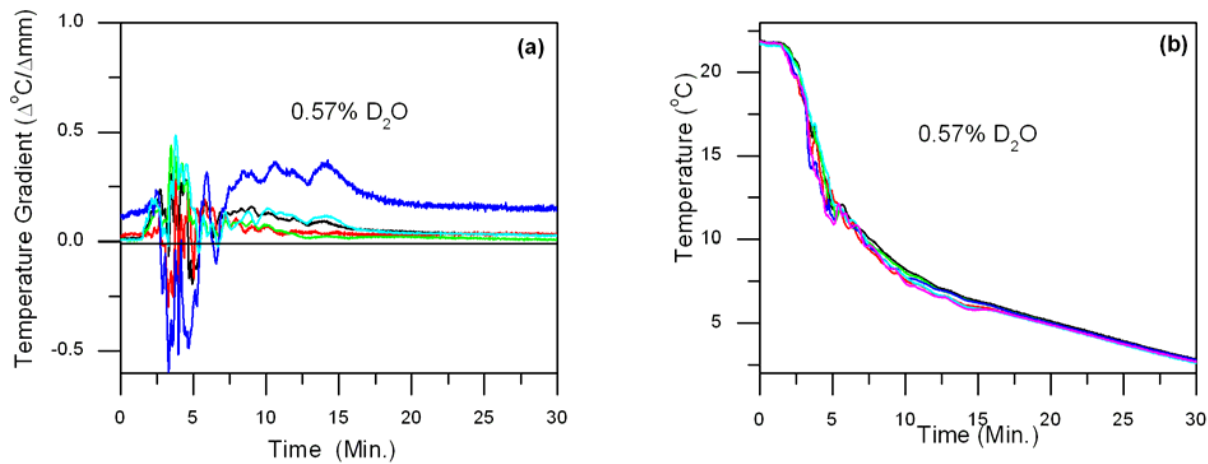

**Fig. 5S.** At 0.57% unmixed D$_2$O we record a temperature gradient larger than 1 $^o$C. Notice that the nearly flat region in the cooling curve seen in Fig. 2S (b) at about 20 min into no longer appear in cooling curves.

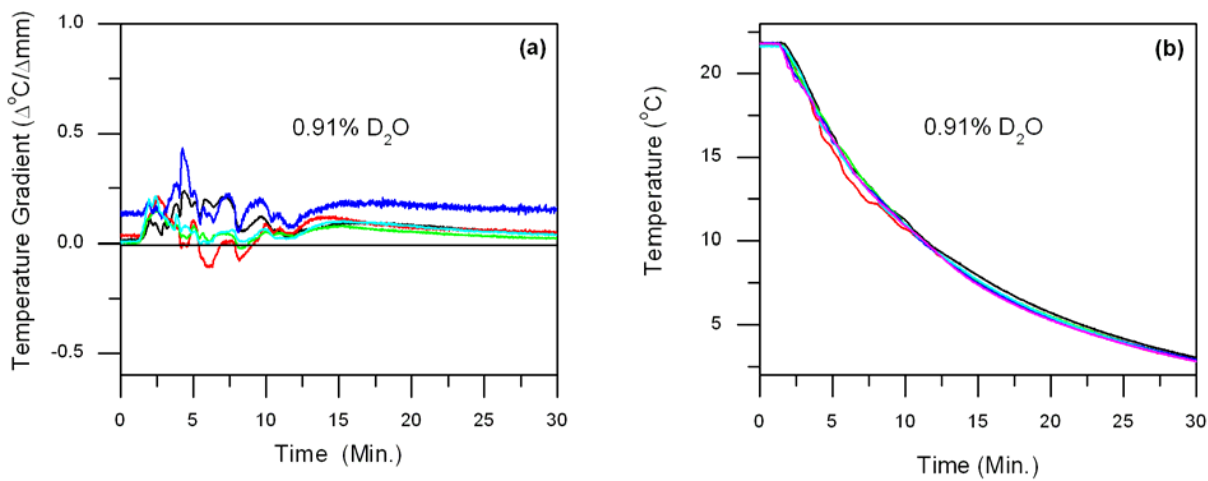

**Fig. 6S.** At 0.91 % D$_2$O the magnitude of the thermal gradient seems to be trending down. We have not yet explored this. Notice the delay in the start of cooling between Fig. 2S (b) and 6S (b).

# Heavy water causes delay in onset of cooling

An example of the thermal fluctuations is shown below in Fig. 7S. Here we show that an increase in $D_2O$ concentration causes the thermal fluctuations to increase. It is also interesting to note that the time it takes the thermocouple registered a change in temperature increases with an increase in concentration. At time t = 0 the container was placed in the ice bath. Notice the increasing delay in response of the thermocouple as we increase the concentration of $D_2O$. The response for 0%, 0.23% and 0.46% added $D_2O$ was 23 sec, 55 sec, and 86 sec respectively.

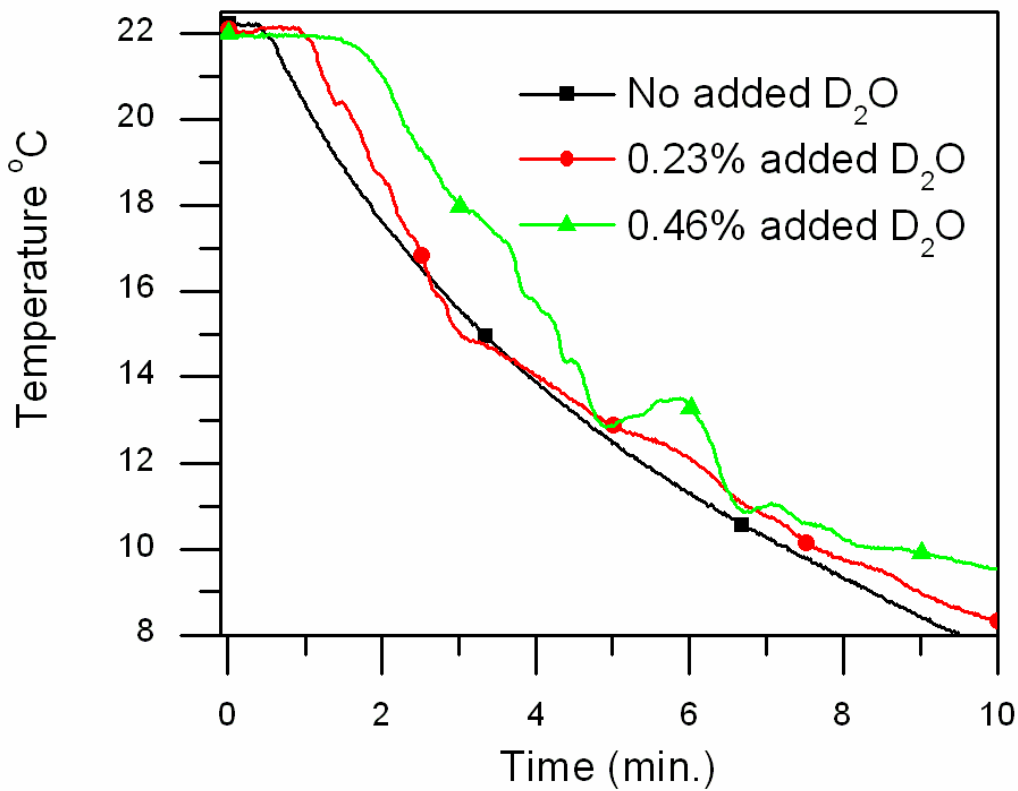

**Fig. 7S.** Here we show how the delay in the loss of heat from the center of a container is affected by adding heavy water to regular water. In addition to delaying the movement of heat from the center of the container, thermal fluctuations are also produced in water with added unmixed $D_2O$. Thermal fluctuations were not produced in water when the $D_2O$ was mixed or not added (the black trace). We show only the first 10 minutes in order to highlight the observed time delay during the first 2 minutes of the cool down.

## Time to self mixing

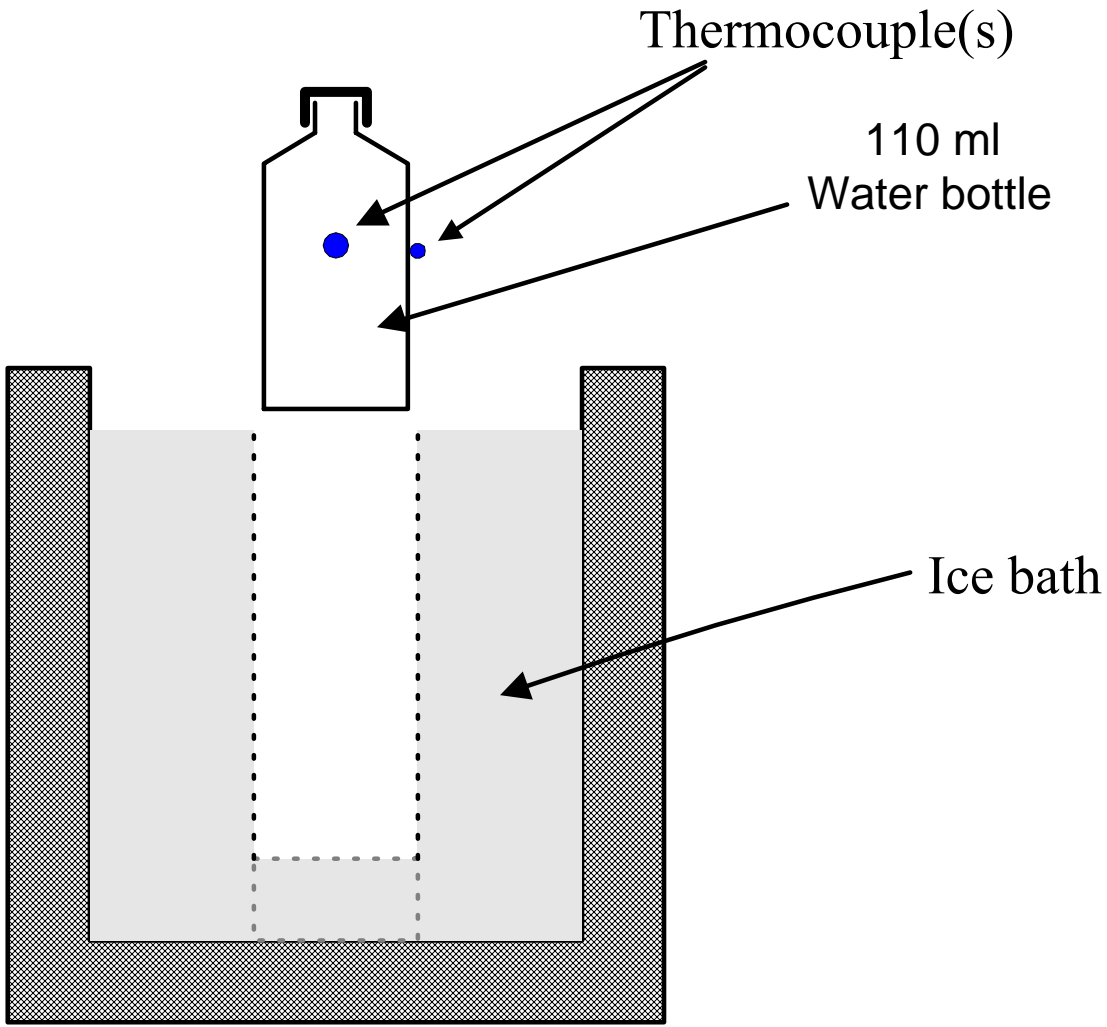

**Fig. 8S.** The schematic of the setup used to remove heat from all surfaces of a 110 ml glass bottle. With a thermocouple attached to the outside of the bottle the time it took heat to flow from the center of the bottle was determined.

Using the setup shown in Fig. 8S, we attempted to answer the question of how long will it take for $D_2O$ to become thoroughly self mixed in a stationary column of ordinary water? The following procedures were used: We sat 20 glass bottles in a controlled environment each filled with 110 ml of ordinary deionized water. To 14 we added 10 drops of $D_2O$ (~0.23% by volume) and we let 11 of the 14 set undisturbed at room temperature for a period of from 12 to 100 hours before cooling. The 3 remaining bottles were rotated 180º several times and immediately cooled. Six additional bottles of ordinary water with no $D_2O$ added were also immediately cooled. These nine bottles (3 mixed $D_2O$ and 6 $H_2O$) serve as a reference to determine when added heavy water was completely "mixed" with ordinary water. Just prior to cool a thermocouple was inserted into the center of the bottle and the container was then carefully placed in an ice bath to cool as shown in Fig. 8S. The results are shown Fig. 9S.

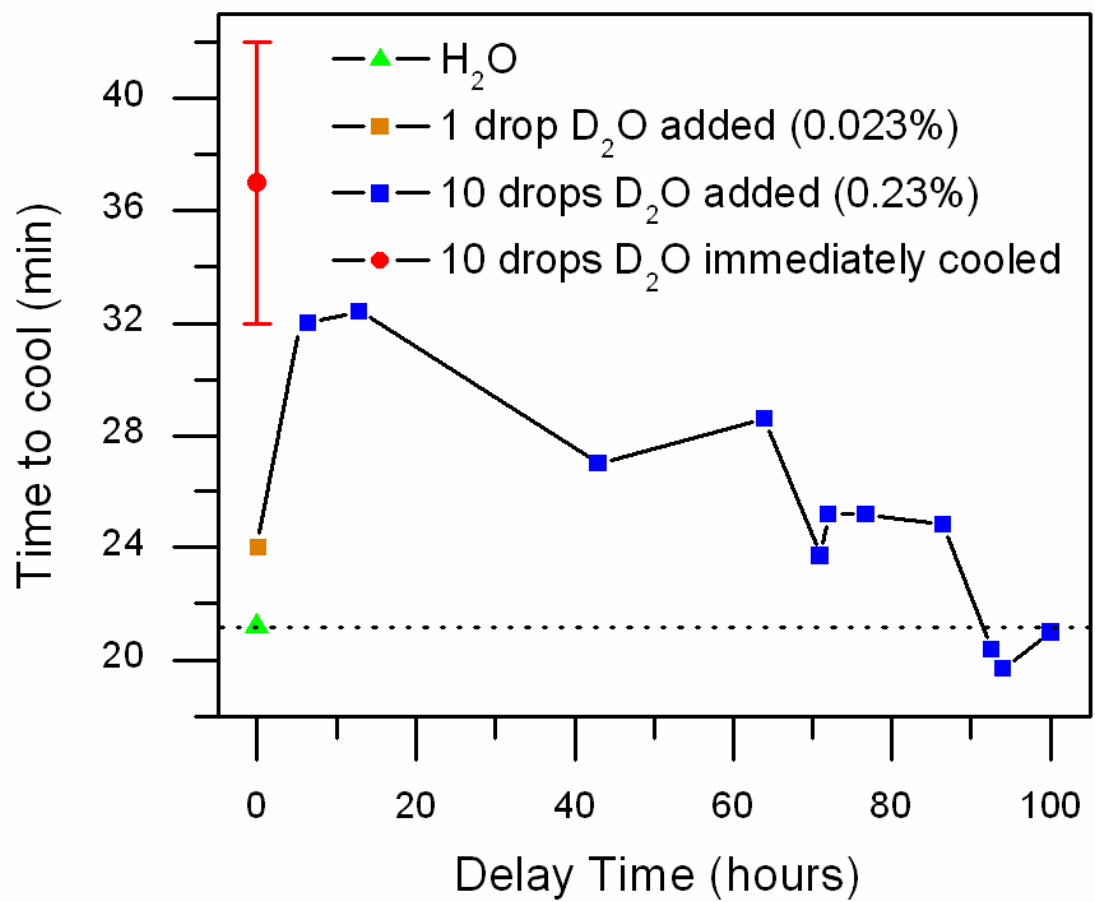

**Fig. 9S.** The delay time is the elapsed time after the addition of D2O. The green triangle is the result of cooling 5 bottles of water with no added heavy water **(error bars are smaller than symbol)**. The red circle, with large error bars, is an example of the size of the error that may be caused by partial mixing; when a bottle is moved from the storage site, when the thermocouple inserted or when it is placed in the ice bath or other wise disturbed. The blue squares, with the connecting lines, show how self-mixing progresses with shelf life, with each point representing one cooling cycle. After about 90 – 100 hours in a controlled environment, at ~22°C, mixing is complete.

## Reproducibility of cooling curves

To investigate the reproducibility of this phenomenon, 5 runs of $H_2O$ and 5 runs with 0.023% $D_2O$ were conducted. For the $H_2O$ the cooling curves were identical up until the sudden drop in temperature as the temperature passes through the maximum density of water at ~4°C, as shown in Fig. 10S(a). Fig. 10S(b) represents the curves where 0.23% $D_2O$ was added to $H_2O$ and shows a much greater difference between runs before the temperature fell below 4°C.

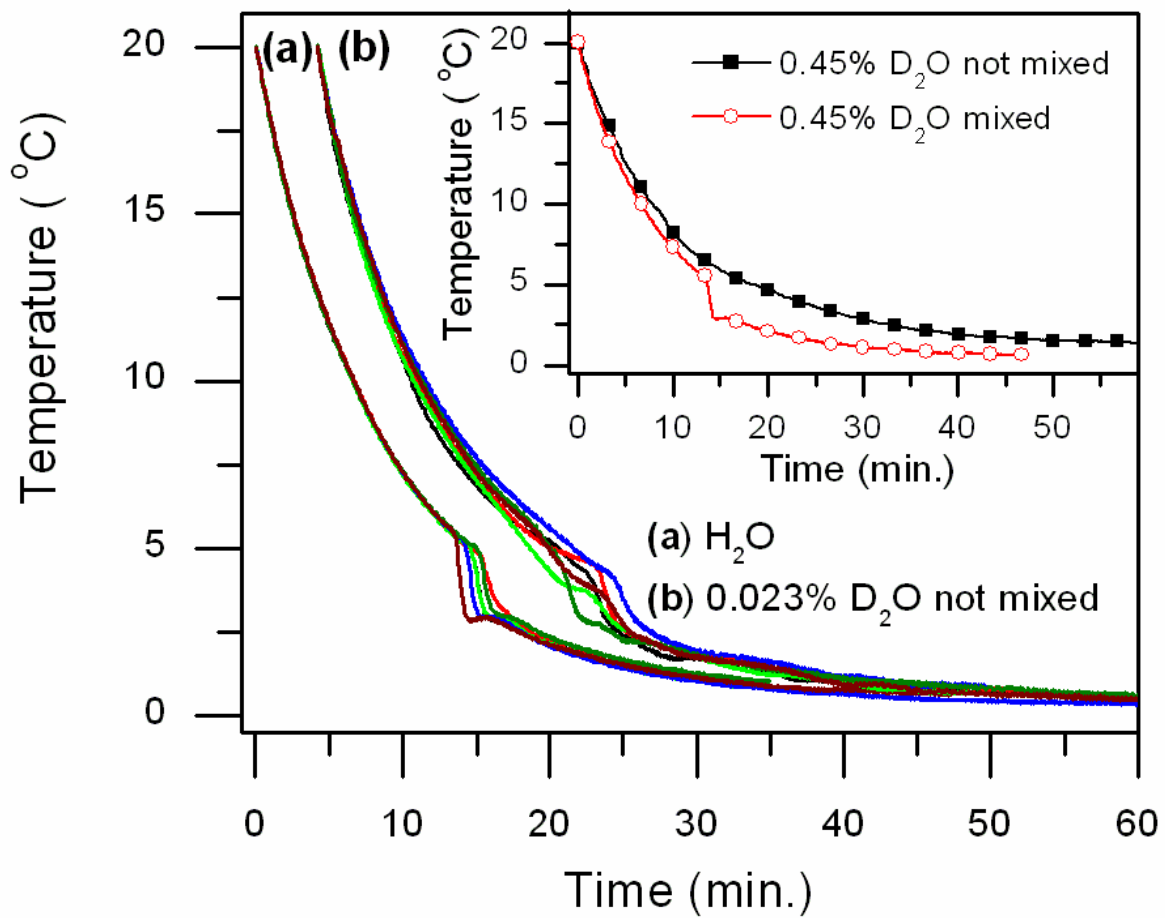

**Fig. 10S.** The reproducibility of the cooling curve for water with no heavy water added is shown in **(a)**. In **(b)** we show the effect of adding just 0.023% heavy water to regular water. The inset is an example of how the cooling curve changes when the amount of heavy water was increased to 0.45%. The sudden drop in temperature below 5 °C is associated with the density maximum of water at 3.98°C.[7]

# Change in temperature causes thermal fluctuations

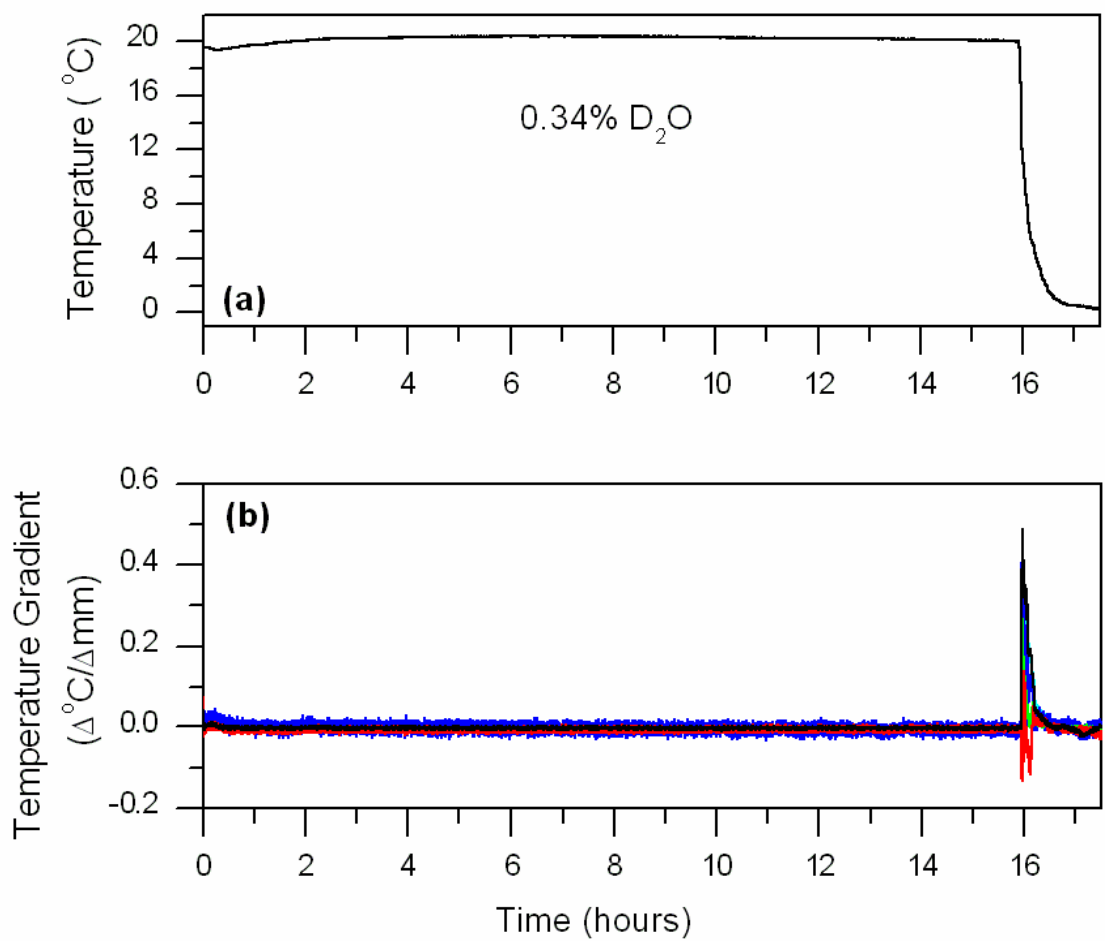

**Fig. 11S.** Temperature versus time for 0.34% $D_2O$ added to light water to show that the effect is seen only during a change in temperature and not when the container sets in a controlled environment at room temperature for ~16 hours. For water with no added $D_2O$ or if the $D_2O$ was mixed little or no temperature gradient is observed when the water began to cool. The setup shown in Fig. 8S was used to collect this data.

- 15 -

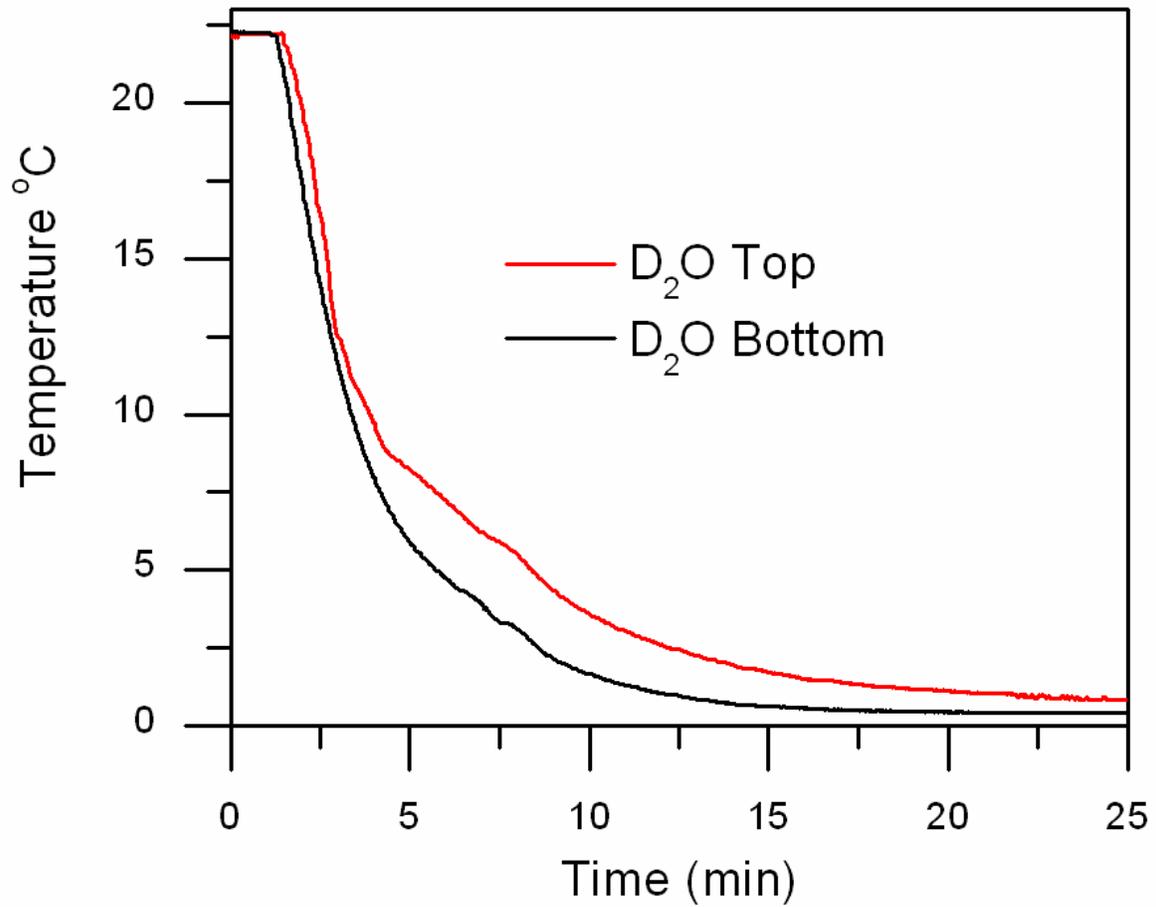

**Fig. 12S.** Here we show cooling curves for water with added unmixed $D_2O$. Notice the dramatic effect adding 0.023% at the top vs. the bottom.

$D_2O$ added to $H_2O$

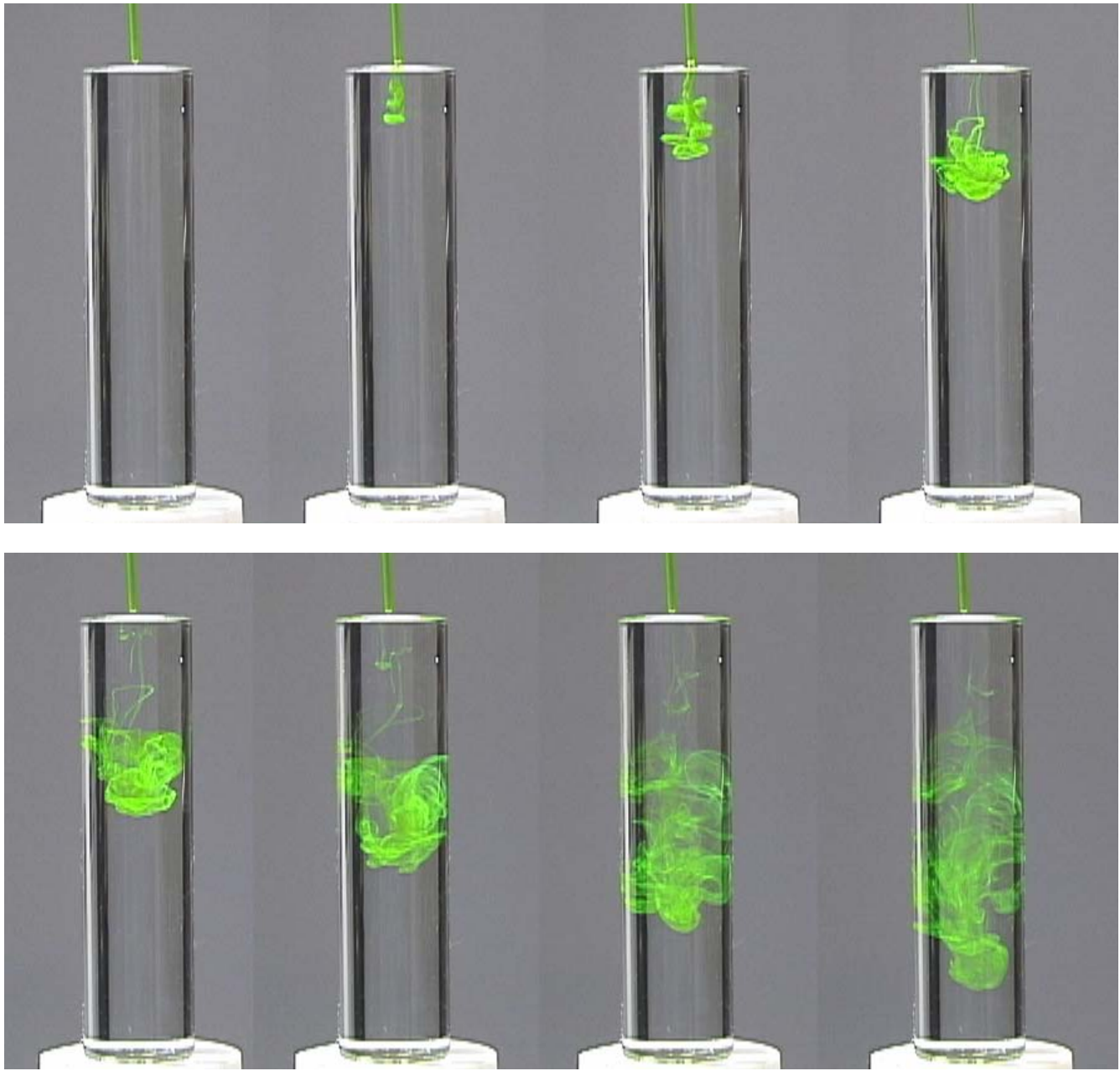

**Fig. 13S.** Here we show several snapshots of a container of water after a drop of fluorescein stained $D_2O$ was added. This sequence of photos was taken in about **5 seconds**. Two drops of heavy water are captured sinking to the bottom. 0.025 ml of fluorescein was added to 2ml of heavy water as a tracer.

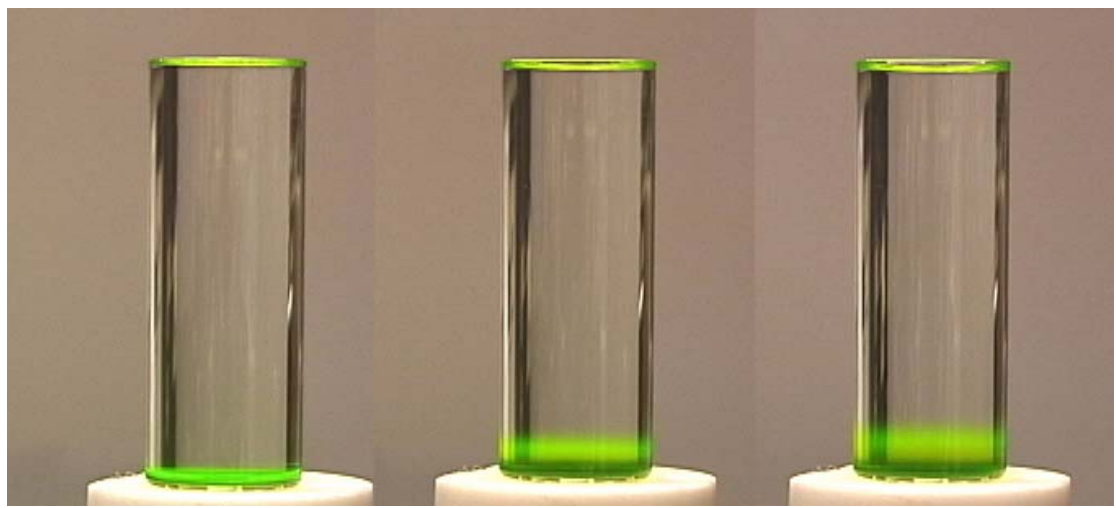

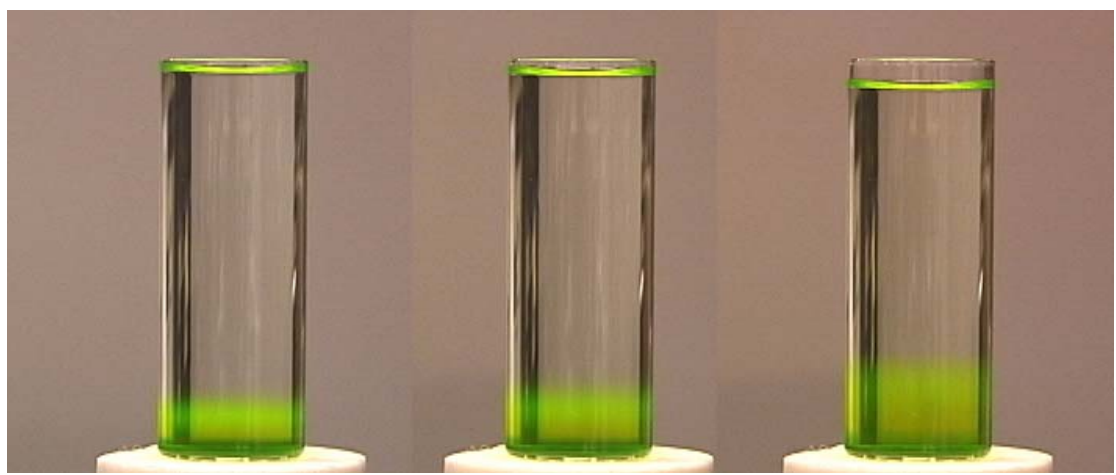

**Fig. 14S.** Here we show several snapshots of a container of water after fluorescein stained D$_2$O was added to the bottom. The last snapshot was taken **24 hours** after the D$_2$O was added. Notice the lost of water due to evaporation. The green color at the top is due to reflection from the bottom.